\documentclass[bibnotes, aps, onecolumn, superscriptaddress]{revtex4}%
\usepackage{epsfig,dsfont,amssymb,amsmath,amsthm,amsfonts,amsbsy,mathrsfs}
\usepackage{graphicx}
\usepackage{lipsum}
\usepackage{bm}

\begin{document}
\title{Synchronous oscillations locked on classical energy levels by two cooperating drives}
\author{Bing He}
\altaffiliation{Contributed equally to this work}
\affiliation{Center for Quantum Optics and Quantum Information,Universidad Mayor, Camino La Pir\'{a}mide 5750, Huechuraba, Chile}
\author{Qing Lin}
\altaffiliation{Contributed equally to this work}
\affiliation{Fujian Key Laboratory of Light Propagation and Transformation, College 
of Information Science and Engineering, Huaqiao University, Xiamen 361021, China}
\author{Miguel Orszag}
\affiliation{Center for Quantum Optics and Quantum Information,Universidad Mayor, Camino La Pir\'{a}mide 5750, Huechuraba, Chile}
\affiliation{Instituto de F\'{i}sica, Pontificia Universidad Cat\'{o}lica de Chile, Casilla 306, Santiago, Chile}
\author{Min Xiao}
\affiliation{Department of Physics, University of Arkansas, Fayetteville, AR 72701, USA}

\begin{abstract}
It is intuitively imagined that the energy of a classical object always takes continues values and can hardly be confined to discrete ones like the energy levels of microscopic systems. Here, we demonstrate that such classical energy levels against intuition 
can be created through a previously unknown synchronization process for nonlinearly coupled macroscopic oscillators driven by two equally strong fields. Given the properly matched frequencies of the two drive fields, the amplitude and phase of an oscillator will be frozen on one of a series of determined trajectories like energy levels, and the phenomenon exists for whatever drive intensity beyond a threshold. Interestingly, the oscillator's motion can be highly sensitive to its initial condition but, unlike the aperiodicity in chaotic motion, it will nonetheless end up on such fixed energy levels. Upon reaching the stability, however, the oscillations on the energy levels are robust against noisy perturbation.
\end{abstract}

\maketitle

\section{Introduction}

Driving nonlinear systems can give rise to interesting phenomena. One category of 
these phenomena is dynamical synchronization \cite{synchronization1, synchronization2, synchronization3, synchronization4}, which has been studied since the time of C. Huygens \cite{origin}.  The frequencies and phases of multiple oscillators can be synchronized under weak mutual interaction, to exhibit the behaviors such as the coordinated flashes of fireflies \cite{firefly} and the injection locking of a laser array to increase output power \cite{laser}. Synchronization is accompanied by mode locking. When it is synchronized by a periodic force of constant amplitude, a nonlinear oscillator will be locked to a number of frequencies known as the devil's staircase. A display of the phenomenon in real physical system is the voltage-current relation called Shapiro steps for a Josephson junction in AC field \cite{step, review1}. Accordingly, one may ask the question---whether the amplitude of an oscillation can also be locked to a number of fixed values at the same time? For example, by locking the amplitude $A$ of a mechanical oscillation $X_m(t)=A\sin(\omega_m t)$ with the frequency $\omega_m$, the energy ${\cal E}_m=\frac{1}{2}(X_m^2+P_m^2)$ of the mechanical oscillator determined by its displacement $X_m(t)$ and momentum $P_m(t)$ will locate on a number of levels corresponding to the locked discrete values $A_n$ ($n\geq 1$), as if its quantization were realized only by the means of classical physics. For a macroscopic object it is against intuition to conceive the possible existence of its discrete energy levels.

We show that the energy levels like the above mentioned can be created for a macroscopic object through a process of synchronization by two different drives. 
It is through a general model in Fig. 1A, which can be experimentally implemented by driving a cavity field pressurizing 
on a mechanical oscillator with two coherent fields having their specific frequencies. 
The previous researches on the similar doubly driven optomechanical systems always concern one strong and one weak field \cite{oms}, such as in the optomechanically induced transparency \cite{eit1,eit2, eit3} and the optomechanical chaos \cite{chaos-theory,chaos}, in addition to the study of mechanical squeezing induced by two drives of different amplitudes \cite{sq}. Instead, the phenomena illustrated below emerge under two drives with equal amplitudes $E_1$ and $E_2$. General nonlinear dynamics due to two or more different external drives has not been well explored thus far, except for the stochastic resonance phenomenon involving one noise drive \cite{sr, noise-review}. Among the unexplored phenomena of optomechanical systems under two drives, we focus on those due to one red detuned drive ($\omega_1=\omega_c-\omega_m$) and 
one resonant drive ($\omega_2=\omega_c$). If acting alone, the former achieves the cooling effect of reducing the mechanical fluctuation in thermal environment \cite{oms}. The two drives work together to bring about a type of previously unknown synchronization to two coupled oscillators that model the system. Such synchronization simultaneously locks the oscillation frequency components and their phases for the two oscillators, as well as the amplitudes for one of them to form the energy levels. 

The real-time evolution of the system toward the energy levels also exhibits previously unknown dynamical behaviors such as the sensitivity to drive field amplitudes and initial conditions. One particularly interesting of them is that the system can evolve to a different energy level if its initial condition changes a little bit. This type of sensitivity differs from the well known character of chaos in that, instead of 
having exponentially growing difference, the final states due to slightly different initial conditions are always on two energy levels with fixed difference. In contrast to the transient period when the initial condition and noises can affect the evolution course, the finally stabilized oscillation on an energy level is rather robust against the external perturbations, so that these energy levels can be observed. 

\begin{figure}
\includegraphics[width=10.8cm]{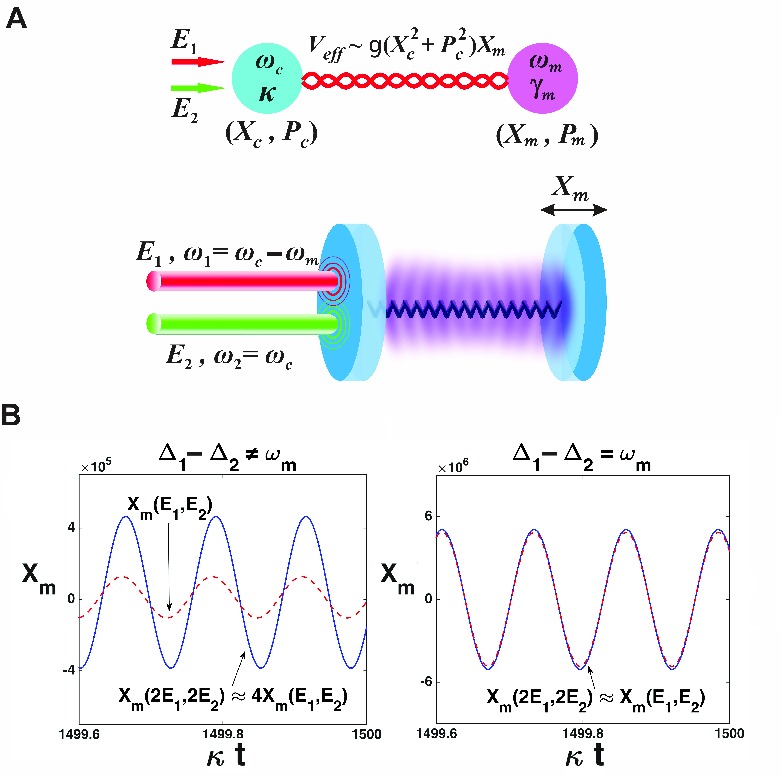}
\centering
\caption{{\bf Model of two nonlinearly coupled oscillators under two external drives.} ({\bf A}) The setup of two drives on a cavity with a fixed mirror 
and a movable mirror (the mechanical oscillator) connected by a spring. This system exemplifies a general model of two oscillators with the intrinsic frequencies $\omega_c$ and $\omega_m$, and the damping rates $\kappa$ and $\gamma_m$ ($\gamma_m \ll \kappa$ as in \cite{oms}), respectively. They are coupled by the interaction potential 
$V_{eff}$, which is realized by modifying the cavity frequency $\omega_c$ with the displacement $X_m$ much less than the cavity length. ({\bf B}) The stabilized $X_m(t)$ of the mechanical oscillator linearly responses to the increase of the drive amplitudes, when their frequencies do not match ($\Delta_1=1.002\omega_m$ and $\Delta_2=0$), but $X_m(t)$ becomes frozen under the condition $\Delta_1=\omega_m$ and $\Delta_2=0$. The relative parameters $\omega_m=50 \kappa$, $g/\sqrt{2}=10^{-5} \kappa$, $\gamma_m=10^{-5}\kappa$ for the system, as well as $E_{1(2)}=2.5\times 10^5\kappa$, are used.}
\label{}
\end{figure} 

\section{Emergence of mode locking phenomenon}

In terms of the two coordinates $X_c$ and $P_c$ of one oscillator in its phase space (corresponding to two perpendicular quadratures of the cavity field), together with those of the other oscillator (corresponding to the displacement $X_m$ and momentum $P_m$ of the mechanical oscillator), the dynamical equations of the system in Fig. 1A read 
\begin{align}
\dot{X}_c  &  = -\kappa X_c-gX_mP_c+\sqrt{2}E_1\cos(\Delta_1 t)+\sqrt{2}E_2\cos(\Delta_2 t),\nonumber\\
\dot{P}_c &= -\kappa P_c+gX_mX_c+\sqrt{2}E_1\sin(\Delta_1 t)+\sqrt{2}E_2\sin(\Delta_2 t), \nonumber\\
\dot{X}_m &=\omega_m P_m, \nonumber\\
\dot{P}_m &=-\omega_m X_m-\gamma_m P_m+\frac{\sqrt{2}}{4}g(X_c^2+P_c^2)
\end{align}
in the observation system rotating at the frequency $\omega_c$, where $\Delta_{1(2)}=\omega_c-\omega_{1(2)}$.
A realistic system always has a very small coupling constant $g$ for the quadratic terms in the equations, which, by appearance, simply correct the linear solution at $g=0$. However, to a driven system like this, the nonlinear terms can govern the system dynamics. One such example is given in Fig. 1B---with a small deviation from $\Delta_1=\omega_m$ and $\Delta_2=0$, the displacement $X_m(t)$ responds linearly to the drive amplitudes, but the drive frequency match locks the amplitude, frequency and phase of $X_m(t)$ totally, so that the stabilized oscillations become almost the same.

\begin{figure}
\includegraphics[width=16.8cm]{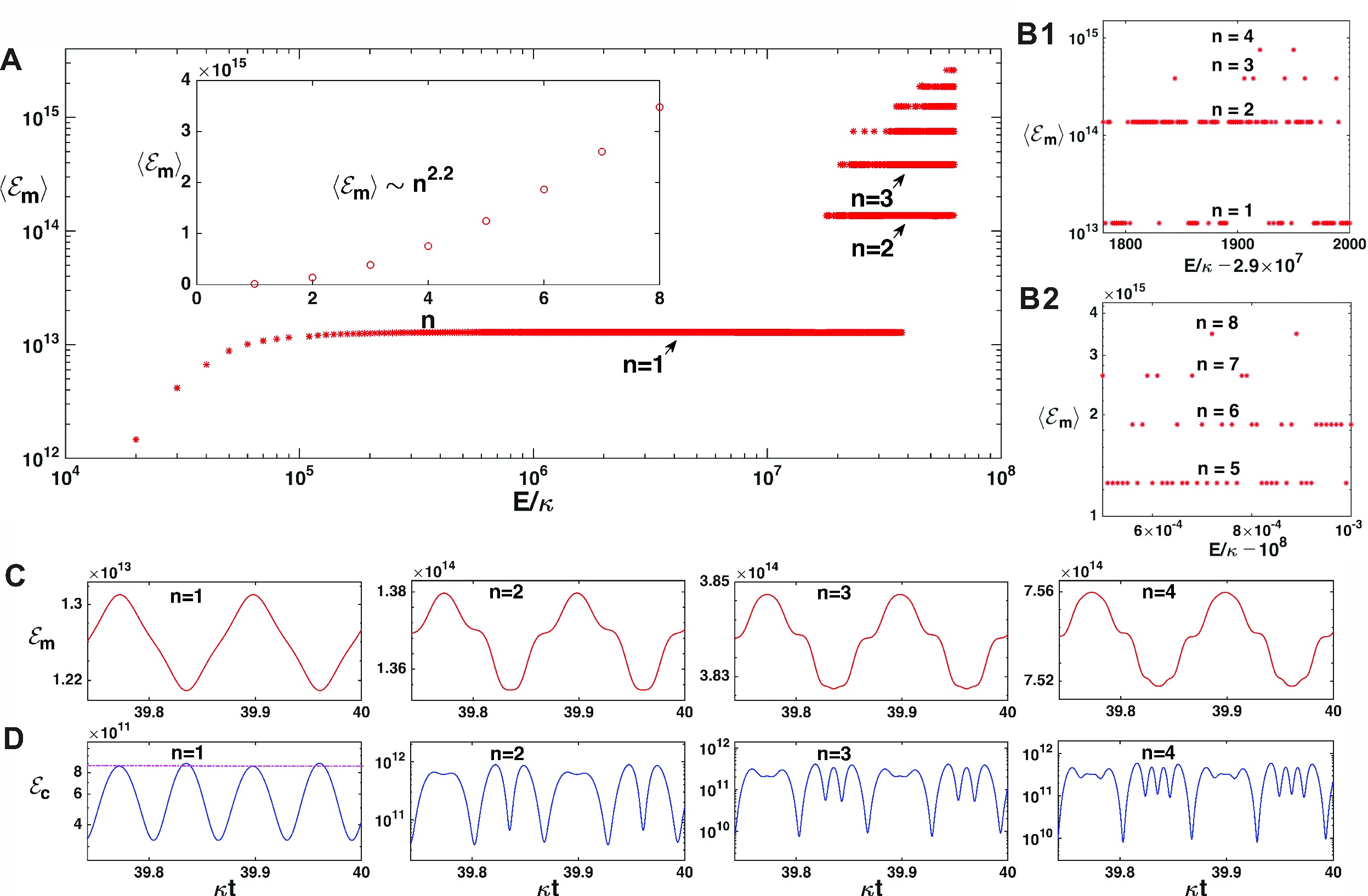}
\centering
\caption{{\bf Mechanical energy levels and associated oscillation patterns.} ({\bf A}) The beginning energy levels demonstrated with their relations to the 
dimensionless drive amplitude up to $E/\kappa=6.3\times 10^7$, in terms of 
the logarithmic scales. All values of $\langle {\cal E}_m \rangle$ displayed are 
the dynamically stabilized ones. These energy levels go up like a quasi parabola as shown in the inset. The system parameters are the same as those in Fig. 1B. A higher (lower) nonlinear coupling $g$ gives the decreased (increased) 
energies on the levels. ({\bf B1}) One section along the horizontal axis of Fig. 2A, viewed with the scale in the order of $10^3$. The distribution of the stabilized average mechanical energy along the horizontal axis is irregular, but their values on the vertical axis are completely fixed. ({\bf B2}) The view of another range starting from $E/\kappa=10^8$ with the scale of $10^{-4}$ (the logarithmic scale on the vertical axis appears uneven). In this range a level transition takes place with $\delta E=10^{-12}E$. ({\bf C}) and ({\bf D}) The one-to-one correspondence between the stabilized mechanical oscillations and cavity oscillations. Here we use a dash line to make the peak of $n=1$ distinct.}
\end{figure} 

\section{Properties of energy levels}

In an optomechanical system that realizes the mode locking phenomenon, the mechanical oscillator with the frequency $\omega_m$ is under the pressure of the cavity sidebands with various frequencies, among which only the one with the frequency $\omega_m$ contributes significantly to the mechanical oscillation due to the resonance effect. 
Then the stabilized mechanical motion can be approximated by a single-frequency oscillation 
\begin{eqnarray}
X_m(t)&=&A \sin(\omega_m t+\phi)+d,
\label{mechanic}
\end{eqnarray}
where $A$ and $d$ are the amplitude and pure displacement of 
the oscillator, respectively; see more details in Methods. The mode locking phenomenon illustrated in Fig. 1B means that the oscillation amplitude $A$ becomes frozen to a range of increased drive amplitude $E$ (this notation stands for $E_1=E_2$), while the accompanying displacement $d$ changes slightly with $E$. In this situation the amplitude has the discrete values $A_n$, and the mechanical energy ${\cal E}_m(t)=\frac{1}{2}(X_m^2(t)+P_m^2(t))$, which is a half of the squared radius of the oscillator's position in its phase space, takes the form 
\begin{eqnarray}
{\cal E}_m(t)
=\frac{1}{2}A_n^2+\frac{1}{2}d_n^2+A_nd_n\sin (\omega_m t+\phi_n),
\label{levels}
\end{eqnarray}
where $d_n\ll A_n$. The time average energy of the mechanical oscillation for each $A_n$ displays the energy levels $\langle {\cal E}_m\rangle(n)\approx \frac{1}{2}A_n^2$, though the total energy of the system including the parts of cavity field and nonlinear coupling is still continuous with $E$. Fig. 1B is equivalent to how the first energy level is created: once the two drives satisfy $\Delta_1=\omega_m$ and $\Delta_2=0$, 
the average mechanical energy $\langle {\cal E}_m\rangle(1)$ on the first level will change very slightly over a considerable range of the drive amplitude $E$. Moreover, on the energy level, the oscillations due to different $E$ take an exactly 
same pace by locking the phase $\phi_1$.

To the abstract model in Fig. 1A, the dimensionless amplitude $E/\kappa$ can be arbitrarily high and, for the real optomechanical systems, sufficiently high $E/\kappa$ is realizable with a cavity of high quality factor. The general distribution of the stabilized energy $\langle {\cal E}_m \rangle$ with the drive amplitude $E$ is shown in Fig. 2A, where it is locked on a series of levels with large differences. Using the system parameters in Fig. 1B, one sees that, around the 
threshold amplitude $E\approx 5\times 10^5 \kappa$, the illustrated system undergoes a dynamical transition from the linear response regime to the first energy level. Over the amplitude $E\approx 2.5\times 10^7 \kappa$, another dynamical transition takes place to have the second energy level emerging and seemingly overlapped with the first one. However, after magnifying the scales on the horizontal axis as in Figs. B1 and B2, there is still one-to-one correspondence between $\langle {\cal E}_m \rangle$ and $E$ in this regime, where the discrete values  $\langle {\cal E}_m \rangle$ of the energy level distributes irregularly along the horizontal axis. This phenomenon will be discussed below. The ``quantized" mechanical energy on the displayed levels satisfies a power law $\langle {\cal E}_m\rangle (n)\sim n^{2.2}$.

The stabilized oscillations after the system has evolved to the energy levels can be found directly with the numerical simulations based on the nonlinear dynamical equations, Eq. (1). Their patterns shown in Fig. 2C are dominated by the frequency component of $\omega_m$, displaying the invariant contours for the oscillations on all levels, 
though one more harmonic component appears after going up one level (the number of the tiny twists on the curves ${\cal E}_m(t)$ or those of $X_m(t)$ and $P_m(t)$ increases in this way). This reflects the fact that the contribution of the high harmonic components 
$n\omega_m$ ($n\geq 2$) to the mechanical energy ${\cal E}_m$ is negligible, being consistent with the single-mode approximation in Eq. (2). 
The amplitudes of the oscillation contours in Fig. 2C are equal to $A_nd_n$. 
 
Corresponding to each energy level, the stabilized oscillation of the cavity energy ${\cal E}_c(t)=\frac{1}{2}(X_c^2(t)+P_c^2(t))$ or the energy of the oscillator directly under the two external drives has a fixed spectrum as one of the invariant patterns illustrated in Fig. 2D. For all different drive amplitudes $E$ leading to the same energy level, the cavity oscillation patterns only differ by their amplitudes proportional 
to $E$. The level, on which the mechanical oscillator is, can thus be known from the peak number in a half period of the cavity field intensity, a unique spectrum of the cavity field that can be detected. The phases of the mechanical spectrum, from the base frequency $\omega_m$ to the numerically discovered high harmonic components, are completely synchronized with those of the cavity oscillation (in the form of $n:m$ synchronization \cite{synchronization2, synchronization4} for the frequency components, where $m$ and $n$ are integers). Such synchronization between the two oscillators of the abstract model (Fig. 1A) is realized under a pair of cooling and resonant fields, in addition to another type of synchronization on each energy level---the mechanical oscillation phases induced by different drive amplitudes leading to the same energy level, as well as those of the corresponding cavity oscillations, are synchronized too; by the single-mode approximation the phase $\phi_n$ in Eq. (\ref{levels}) for each level is identical. Phase dynamics \cite{phase} is the primary concern in synchronization problems, including those in chaotic systems \cite{chaos0,chaos1, chaos2} and systems operating in quantum regime \cite{s1,s2,s3}. The uniqueness in the current problem is a simultaneous phase locking on all entrained frequency components of the two oscillators that model the system, rather than on a couple of frequency components only.

\begin{figure}
\centering
\includegraphics[width=5.8cm]{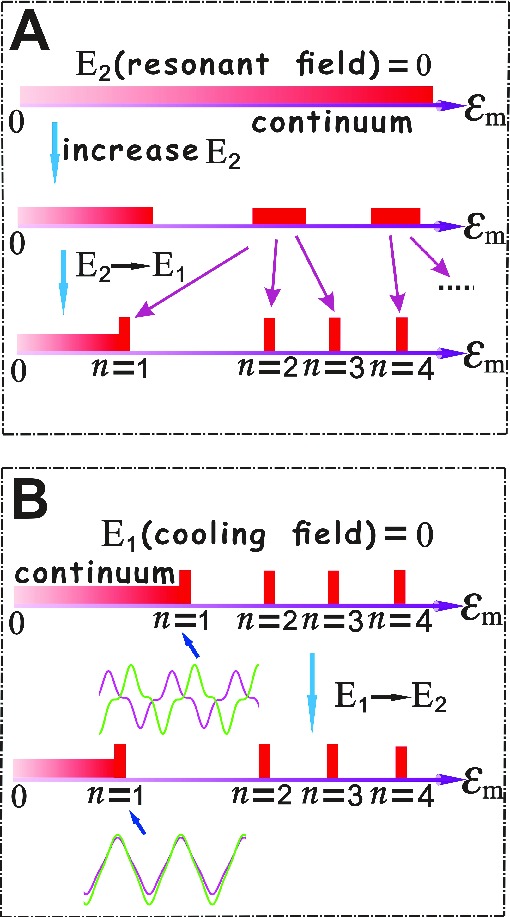}
\caption{{\bf Functions of two different drive fields.} ({\bf A}) From the continuum spectrum due to the sole action of the cooling field, the energy bands and energy levels appear after gradually strengthening the resonant field. ({\bf B}) The energy levels generated by the resonant field alone are lowered by the increased cooling field amplitude. Here the oscillations on an energy level are synchronized after adding the cooling field. In the illustrated example with the level $n=1$, the high harmonic components are also filtered out in the process.}
\end{figure} 

\section{Respective roles of drive fields}

The mechanical oscillations locked on the energy levels can be better understood by starting one of the drives individually; see the illustrations in Fig. 3. From the detailed examples in Figs. D1 and D2 in Appendix D, one finds the functions of the two different drives. A strong resonant field gives rise to significant mechanical displacement and gain effect. Together with the nonlinear saturation, these effects make the stabilized $\langle {\cal E}_m\rangle$ jump by discrete steps together with the resonant drive being continuously enhanced beyond a certain value. After adding such strong resonant field, the continuum spectrum due to a sole cooling field will be split into energy bands and energy levels as in Fig. 3A. The action of a single resonant field with sufficiently high amplitude brings about nonlinear dynamical behavior, in contrast to the regime of weaker drives where the nonlinearity should appear under well matched frequencies of two drives (see Fig. 1B). On the other hand, the effects of the cooling field as illustrated in Fig. 3B are to lower the discrete energy levels created by the resonant field and synchronize the oscillations on a certain level to the same phase. 
The top of the continuum part of the spectrum due to a resonant field alone is lowered by the cooling field so that the first level can be realized with $E\sim 10^5 \kappa$ for our illustrated system. In terms of the effective cooling intensity $J=g E/\omega_m$ \cite{coupling}, the level $n=1$ in our example exists at $J\approx 0.1$, which is experimentally achievable by the current optomechanical systems. The coexisting cooling and resonant fields with equal amplitudes balance the classical energy levels to the fixed positions. 

\section{Transition between energy levels}

Above the drive amplitude leading to the levels $n\geq 2$, the transient processes of evolving to the energy levels are complicated. As it has been shown in Fig. 2B1, a magnified view of one section along the horizontal axis of Fig. 2A that starts from $E=2.9 \times 10^7 \kappa$, the system would go to another level whenever $E$ is shifted to $E+\delta E$ with $\delta E\sim 10^{-5}E$. 
When viewed with a large scale of $E/\kappa$, the energy levels in Fig. 2A thus appear to overlap from the starting point of the second level, unlike the step-by-step devil's staircase \cite{synchronization2, step} in other synchronization phenomena. The transition to other levels occurs with even less change of $E$, given a still larger $E$ as in Fig. 2B2. Here the level transition means the evolution to different levels from the same initial condition rather than a direct jump between the levels, and whether the transition due to $\delta E$ is to a lower or upper one follows a random pattern. In Figs. 2B1-2B2 the energy $\langle {\cal E}_m\rangle$ distributes irregularly along the horizontal axis, but its values on the vertical axis are nonetheless fixed to those of the energy levels. Such phenomenon mainly comes from the effects of the resonant field, due to which the stabilized oscillation amplitude $A_n$ of the oscillator takes random transition under a slight variation of its drive amplitude, in contrast to the overall tendency that a large increase of the resonant field makes the energy levels be higher.

The sensitivity of an evolution process to drive amplitude provides an unusual example that the straightforward perturbation treatment for the dynamics related to external drives breaks down. From such scenario due to the small changes $\delta E_1=\delta E_2$ taking action since the beginning $t=0$, it is imaginable that a tiny fluctuation in either of two high amplitudes $E_1$ and $E_2$, be it a deterministic one with the regular time function or a stochastic one in the form of noise, would make the system evolve to a different energy level. However, it actually depends on when the fluctuation exists. Fluctuations will influence the evolution of the system only when they act before the system has stabilized. Upon evolving to the stability, the oscillations on the energy levels are rather robust against drive fluctuations. 
This feature differing from the noise-induce transitions between quasi-steady states, which are commonly encountered in nonlinear systems \cite{book}, will be elaborated later.
\begin{figure}
\centering
\includegraphics[width=10.8cm]{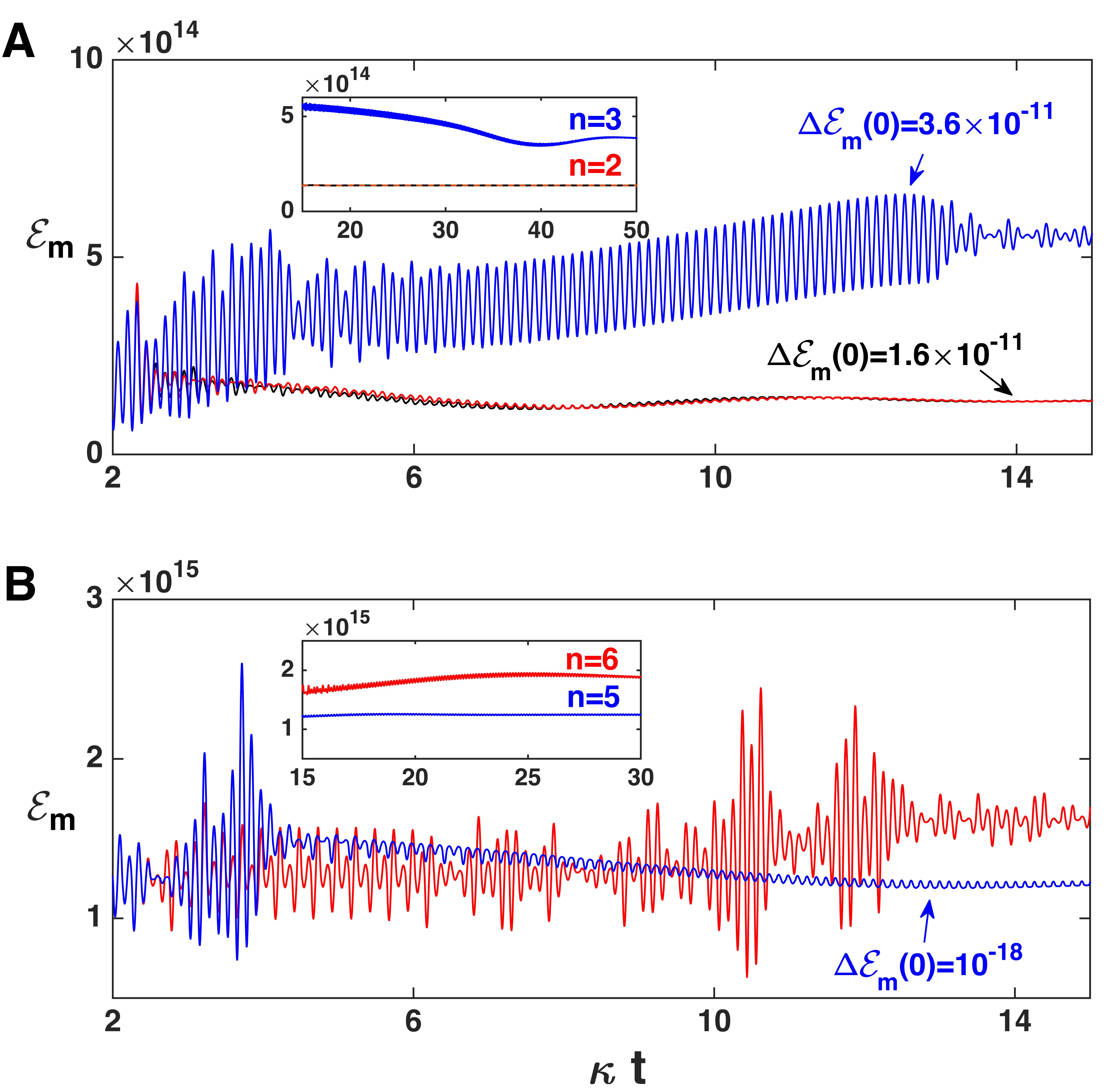}
\caption{{\bf Sensitivity of dynamical evolution to initial conditions.} ({\bf A}) The evolutions of the mechanical energy under the fixed drives with 
$E= 3.5\times 10^{7}\kappa$, but with the differences in the initial conditions, i.e., $(X_m(0),P_m(0))=(0,0)$ for the red, $(X_m(0),P_m(0))=(4\sqrt{2}\times 10^{-6},0)$ for the black curve, and $(X_m(0),P_m(0))=(6\sqrt{2}\times 10^{-6},0)$ for the indigo curve. The indigo one evolves to a different enenrgy level. The inset shows the period of reaching the stability. ({\bf B}) The evolutions of the mechanical energy under the fixed drives with $E=(10^8+6.01 \times 10^{-4})\kappa$, but with a tiny difference in the initial conditions, i.e., $(X_m(0),P_m(0))=(0,0)$ for the red and $(X_m(0),P_m(0))=(0,\sqrt{2}\times 10^{-9})$ for the indigo curve. Note that the logarithmic scale on the vertical axis appears uneven.}
\end{figure} 

\section{Evolution sensitivity to initial conditions}

Among the previously known phenomena of nonlinear dynamics, chaotic motion is typical to have a tiny change of initial condition leading to huge difference in proceeding motion. This character also exists to the model in Fig. 1A, when the drives are strong enough to create the higher energy levels. In Fig. 4, the evolution trajectories of the energy ${\cal E}_m(t)$, which are determined by the corresponding evolutions of $X_m(t)$ and $P_m(t)$, are compared for some differences $\Delta {\cal E}_m(0)=\frac{1}{2}\{(\delta X_m(0))^2+(\delta P_m(0))^2\}$ in the mechanical energy at the beginning. 
The results in Fig. 4A are due to the differences of $\delta X_m(0)$ in the order of $10^{-6}$, one of which changes the evolution course to another energy level. The dependence of the evolution course on the initial condition will become more 
sensitive when the external drive fields are stronger to realize the even higher energy levels. Fig. 4B compares the evolution for an oscillator slightly touched at the beginning ($\delta P_m(0)\sim 10^{-9}$) with that of an unperturbed oscillator. The initial difference of the mechanical energy from that of the untouched oscillator is as small as $\Delta {\cal E}_m(0)=10^{-18}$, but the system will take a transition between the fourth and fifth level with a huge gap in the order of $10^{14}$. On the other hand, 
a larger difference in the initial conditions may cause the final stabilization on more different energy levels, in contrast to the other nonlinear systems in which the synchronization is to have the evolutions due to different initial conditions 
locked to the same stabilized motion \cite{synchronization3}.

More interestingly, this type of sensitivity to initial conditions has nothing to do with chaos. There would be the tendency $\lim_{t\rightarrow\infty}\Delta {\cal E}_m(t)/\Delta {\cal E}_m(0)\sim e^{\lambda t}$ where $\lambda$ is a positive Liapunov exponent, were the mechanical oscillator in a chaotic motion. Instead, the real tendency due to a different initial condition is $\lim_{t\rightarrow\infty}\Delta {\cal E}_m(t)/\Delta {\cal E}_m(0)\sim B\sin(\omega_m t)+D$ where $B$ and $D$ are bounded constants (after choosing a proper reference phase). Once the system parameters are determined, the oscillator will evolve to one of the fixed energy levels, no matter how the initial condition is modified. Since there is one-to-one correspondence between the stabilized mechanical oscillation and the stabilized cavity pattern (see Figs. 2C and 2D), the total difference $\Delta {\cal E}_c(t)+\Delta {\cal E}_m(t)$ in the evolutions of the system from the varied initial conditions is bounded too.
\begin{figure}
\centering
\includegraphics[width=8.8cm]{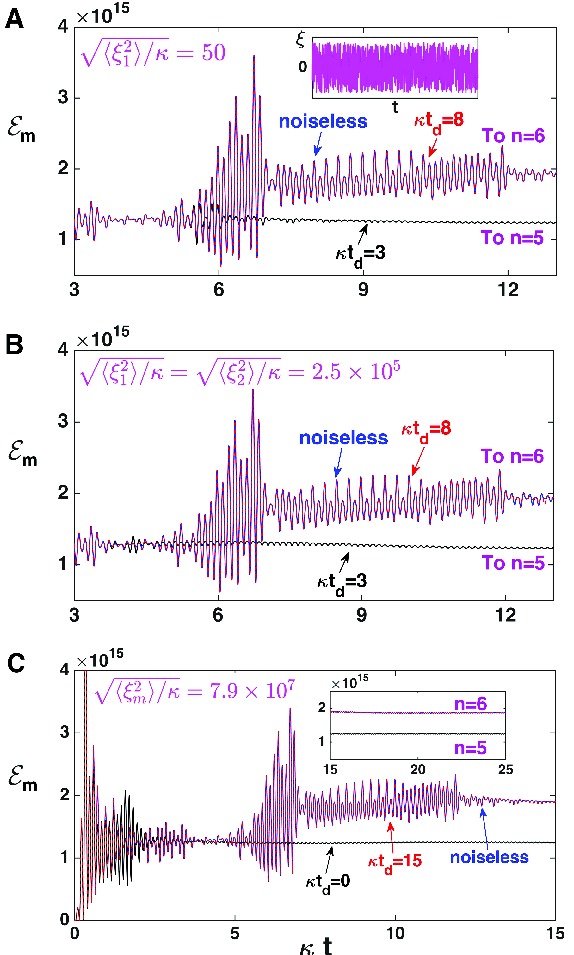}
\caption{{\bf Influence of noisy perturbations.} ({\bf A}) and ({\bf B}) How the evolution courses are affected by the drive amplitude fluctuations in the form 
$H(t- t_d)\xi_{1(2)}(t)$, where $\xi_{1(2)}(t)$ of the respective amplitude $\sqrt{\langle\xi_{1(2)}^2\rangle}$ (the square root variation) is a random function generated with Matlab, as shown in the inset of A. The random signal changes its value for every step size of $\kappa t=1.5\times 10^{-5}$. If the delay time $t_d$ in the Heaviside function is close to the time when the system is stabilized, the red curve due to the noise added later will almost coincide with the blue curve (the noiseless one without the fluctuation of drive amplitudes), even under the one with much higher amplitude. 
({\bf C}) The effect of a random drive $H(t- t_d)\xi_{m}(t)$ on the mechanical oscillator. The noisy perturbation starting at the beginning ($t_d=0$) changes the evolution course (the black curve), but the same noise appearing much 
later will not affect the evolution course (the red curve), which coincides with the course (the blue curve) without such perturbation that simulates the thermal noise.}
\end{figure} 

\section{Robustness of energy levels}

Another important issue is how noisy perturbations affect the classical energy levels created by two external fields with matched frequencies. One kind of noise exists as the fluctuations of the drive field amplitudes, i.e., $E_i$ ($i=1,2$) in Eq. (1) is to be replaced by a more realistic one $E_i+\sqrt{\kappa}\xi_i(t)$, where $\xi_i(t)$ is a stochastic function of time. We here provide a couple of such examples in Fig. 5. First, in Fig. 5A, we consider a random fluctuation with small intensity (its variation $\langle\xi_1^2\rangle$ is low as compared with the intensities of the drives) 
only on the cooling field. 
This fluctuation is added into the cooling drive from two different moments, i.e., its function is multiplied by a Heaviside function $H(\kappa t-\kappa t_d)$ with two different $t_d$. The evolutions of ${\cal E}_m(t)$ after adding the fluctuation are compared with the ideal situation without any noise. It is found that the tiny random fluctuation added before the system reaches the stability will change the evolution course to another energy level. However, if it is added when the system has evolved close to its stability, the same random fluctuation can not affect the motion at all. In Fig. 5B we add much stronger fluctuations to both cooling field and resonant field, and the similar consequences emerge nonetheless. In numerical simulations one can even add a regular fluctuation in the form $\delta E_i\{H(\kappa t-\kappa t_d)-H(\kappa t- \kappa t_s-\kappa t_d)\}$, a square pulse lasting for a period of $t_s$, to the drive amplitudes. Corresponding to the examples in Fig. 5, where the drive amplitude $E$ is in the order of $10^8\kappa$, a sufficiently long pulse ($\kappa t_s>10$) up to $\delta E_i\sim 10^7 \kappa$ will not take any effect if it is input after the system is fully stabilized. These results indicate that the energy levels are rather robust. 

In a thermal environment the mechanical oscillator is under noisy perturbation too.
Then the dynamical equation 
\begin{align}
\dot{P}_m &=-\omega_m X_m-\gamma_m P_m+\frac{\sqrt{2}}{4}g(X_c^2+P_c^2)+\sqrt{\gamma_m}\xi_m(t)
\label{noise}
\end{align}
of the mechanical oscillator has an extra random drive term $\sqrt{\gamma_m}\xi_m(t)$. 
We simulate the evolutions under the action of such random drive in Fig. 5C, where the noise has a sufficiently high magnitude. The results are still similar to those of drive fluctuations: if the noise starts after reaching the stability of the system, the energy levels will be stable forever. Different types of noises (such as colored noises with various spectra) can also be used in Eq. (\ref{noise}), but they do not change the qualitative picture of the system dynamics. The robustness of the stabilized energy levels against noisy perturbation allows their realization in less demanding environment, though the evolution toward a specific energy level is affected by the existing noises.

\section{Discussion}

Our finding reveals the existence of a type of synchronization for two nonlinearly coupled oscillators, which must be realized under two equally strong drives having their frequencies properly matched. Upon entering such synchronization, 
the two oscillators respectively oscillate with a fixed spectrum of entrained frequencies, whose corresponding phases are all synchronized to the same pace. Accompanying the synchronization is the amplitude locking for one of the oscillators, which exhibits a behavior of oscillating on discrete energy levels. Equally increasing the two drive amplitudes beyond a threshold value, one will have the oscillation amplitude of that oscillator to be frozen in spite of the variation of the drive amplitudes over a large range, while its net displacement along the direction of being pushed changes slightly; this is the formation of the first energy level. If the two equal drive amplitudes are still higher than another threshold, the stabilized amplitude of the oscillator can jump to the higher magnitudes by discrete steps, forming the other energy levels. The emergence of the oscillation synchronization locked on the classical energy levels highly relies on the match of the external drive frequencies with those of the coupled oscillators, and then the precise measurement of the intrinsic frequencies of the oscillators can be possible through the observation of such synchronization. 
These phenomena are purely classical because the displacement and 
the momentum of any oscillator are commutative. 

The evolution processes toward the synchronization also exhibit previously unknown dynamical behaviors. A slight fluctuation in drive amplitudes can lead to the transitions between the energy levels with huge difference, but the stabilized energy levels are immune to noisy perturbation from making further transition. This type of sensitivity to perturbation could be applied to detect small changes in environment. Moreover, an evolution course to a higher energy level can become highly sensitive to the initial conditions. However, this differs from chaos because, given any pair of cooperating drives, the system will always be synchronized to a state corresponding to one of the fixed energy levels. As the real physical systems used for example, these dynamical phenomena related to the classical energy levels of a macroscopic object are expected to be observable with the suitable optomechanical systems. 

\appendix

\section{Notations}

The dimensionless variables, $X_c$, $P_c$, $X_m$ and $P_m$, are adopted for the model described by Eq. (1). The conversions of these variables to the real ones are simply by the multiplications of the respective constant factors. 
By the use of these dimensionless variables, the cavity energy 
${\cal E}_c(t)=\frac{1}{2}(X_c^2(t)+P_c^2(t))$ is equivalent to the photon number in the cavity, and the mechanical energy ${\cal E}_m(t)=\frac{1}{2}(X_m^2(t)+P_m^2(t))$ is equivalent to something like a phonon number. All system parameters 
in the dynamical equations have the unit $s^{-1}$ or Hz. The drive amplitude is related to the pump power $P_{1(2)}$ as $E_{1(2)}=\sqrt{\frac{\kappa P_{1(2)}}{\hbar\omega_{1(2)}}}$. For the convenience in the numerical calculations, we use the relative parameters with respect to the cavity damping rate $\kappa$, so that the calculations only involve the dimensionless quantities. For example, the drive amplitudes $E_1$ and $E_2$ are taken as how many times of the parameter $\kappa$.

\section{Relevance of computation precision}

In the physical system that is used for illustrating the concerned synchronization phenomena, two external drives, one cooling field and one resonant field, pump a cavity coupled to a mechanical oscillator. Strong nonlinearity can arise due to the existence of the resonant drive, so, unlike in the cooling of the mechanical oscillator \cite{oms}, the system dynamics cannot be linearized. The classical nonlinear dynamical equations, Eq. (1), are numerically integrable. Another example of numerically approachable nonlinearity in optomechanics is self-sustained oscillation due to one blue-detuned external drive (see, e.g., \cite{self01,self02, self03, self04, self05, self06}), where the stabilized mechanical oscillation amplitude changes continuously with the external drive $E$. For different parameter regimes in our numerical simulations, sufficiently high precision should be chosen at the cost of calculation speed. One tricky point is the simulations involving the higher energy levels ($n\geq 4$), in which a different computation precision will lead to a different energy level being reached in the end. However, any finally stabilized result (given a drive amplitude $E$ beyond the threshold to realize the first energy level) will be magically on one of the fixed energy levels for whatever used precision. This phenomenon happens to reflect the sensitivity of an evolution to the initial condition and the fluctuations at the beginning period, as the different computation precisions are realized by the different iteration step sizes in the algorithms for numerically solving ordinary differential equations. The acceptable results are those not changing with further refined precision. For the lower energy levels (those due to the drive amplitude $E$ up to $5\times 10^7 \kappa$ in our example), there is no such problem caused by computation precision.

\setcounter{figure}{0}
\makeatletter 
\renewcommand{\thefigure}{C\@arabic\c@figure}
\makeatother

\renewcommand{\theequation}{C\arabic{equation}}
\renewcommand{\thetable}{C\arabic{table}}
\setcounter{equation}{0}

\section{Single-mode approximation for locked oscillations}

The single-mode approximation is applied in the interpretation of the displayed energy levels, as in Eqs. (\ref{mechanic}) and (\ref{levels}), while the complete numerical simulations are adopted for finding the locking of oscillations on those levels. 
The approximation means that the stabilized oscillations for the oscillator having its amplitudes locked on the energy levels can take the form
\begin{eqnarray}
X_m(t)&=&A_n \sin(\omega_m t)+d_n
\label{xm}
\end{eqnarray}
by choosing the phase to be zero. Through the numerical simulations based on the nonlinear dynamical equations, Eq. (1), one can find the amplitude $A_n$ of a series of discrete values and the displacement $d_n$ directly, as they can be read from the contours of stabilized oscillation patterns similar to those in Fig. 2C. The examples of their readings are given in the following table:

\begin{table}[h]
\centering
\begin{tabular}
[c]{|c|c|c|c|}\hline
energy level & $A_n$ & $d_n$ & $\langle {\cal E}_m\rangle$ \\\hline
$n=1$ &  $5019460.83$  & $100725.69$  & $1.2869\times 10^{13}$\\\hline
$n=2$ & $16536205.16$ &  $75687.75$ & $1.3694\times 10^{14}$\\\hline
$n=3$ & $27703356.66$ &  $35906.73$ & $3.8385\times 10^{14}$ \\\hline
$n=4$ & $38830599.30$ &  $54012.64$ & $7.5408\times 10^{14}$ \\\hline
$n=5$ & $49948350.03$ &  $46177.52 $ & $1.2476\times 10^{15}$ \\\hline
\end{tabular}
\caption{The oscillation amplitudes $A_n$ and net displacements of the oscillator, as well as the average mechanical energies, on the beginning five energy levels. These values are obtained with the drive amplitudes used in Fig. C1. }
\vspace{0cm}
\end{table}
\noindent On the levels the net displacement $d_n$ has very small variation, to give rise to the width of each energy level.

By plugging the stabilized mechanical oscillation Eq. (\ref{xm}) into Eq. (1), one has the linearized dynamical equations
\begin{align}
\dot{X}_c  &  = -\kappa X_c-g(A_n \sin(\omega_m t)+d_n)P_c
+\underbrace{\sqrt{2}E\cos(\omega_m t)+\sqrt{2}E}_{F_X(t)},\nonumber\\
\dot{P}_c &= -\kappa P_c+g(A_n \sin(\omega_m t)+d_n)X_c
+\underbrace{\sqrt{2}E\sin(\omega_m t)}_{F_P(t)}
\label{linearized}
\end{align}
for the stabilized cavity field quadratures, where $\Delta_1=\omega_m$, $\Delta_2=0$ and $E_1=E_2=E$. The solution to this linear differential equations takes the form:
\begin{align}
&\begin{pmatrix} 
X_c(t)\\
P_c(t)
\end{pmatrix}
=\int_0^t d\tau~{\cal T}\exp\{\int_{\tau}^t dt'
\begin{pmatrix} 
-\kappa &  -g(A_n \sin (\omega_m t')+d_n)\\
g(A_n \sin (\omega_m t')+d_n) & -\kappa
\end{pmatrix}
\} \underbrace{
\begin{pmatrix} 
F_X(\tau)\\
F_P(\tau)
\end{pmatrix}
}_{\vec{\lambda}(\tau)}\nonumber\\
&=\int_0^t d\tau \underbrace{\begin{pmatrix} 
e^{-\kappa (t-\tau)} & 0\\
0 & e^{-\kappa (t-\tau)}
\end{pmatrix}}_{\hat{D}(t,\tau)} \exp\{\int_{\tau}^t dt'
\underbrace{\begin{pmatrix} 
0 &  -g(A_n \sin (\omega_m t')+d_n)\\
g(A_n \sin (\omega_m t')+d_n) & 0
\end{pmatrix}}_{\hat{M}(t')}\}\vec{\lambda}(\tau)\nonumber\\
&=\int_0^t d\tau \hat{D}(t,\tau)\vec{\lambda}(\tau)+\int_0^t d\tau \hat{D}(t,\tau)\int_\tau^t dt'\hat{M}(t')\vec{\lambda}(\tau)+\frac{1}{2!}\int_0^t d\tau 
\hat{D}(t,\tau)\big(\int_\tau^t dt'\hat{M}(t')\big)^2\vec{\lambda}(\tau)\nonumber\\
&+\cdots,
\end{align}
where the time-ordered exponential function in the solution is factorized into the product of two ordinary exponential functions of matrix.

\begin{figure}
\centering
\includegraphics[width=17.5cm]{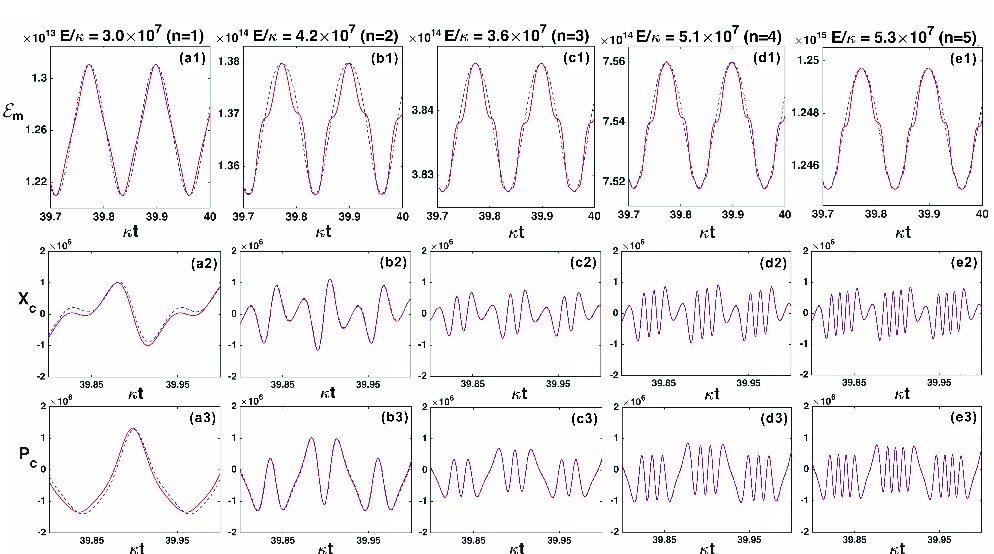}
\caption{{\bf Comparison of the stabilized mechanical energy and associated cavity quadratures predicted with the exact nonlinear dynamical equations and the linearized equations.} The solid curves are obtained with the numerical simulations based on the nonlinear dynamical equations (Eq. (1)), and the dashed curves are from the linearized Eq. (\ref{linearized}). Here, five different drive amplitude values respectively leading to five different energy levels are used for the illustrations from (a1)-(a3) to (e1)-(e3). The drive amplitude $E$ leading to the level $n=2$ is lower than the amplitude that realizes the level $n=3$, as a manifestation of the phenomenon 
illustrated in Figs. 2B1-2B2.}
\end{figure} 

The integrals involving the trigonometry functions in the above equation can be straightforwardly performed to find all Fourier components of the cavity field.
The amplitude $A_n$ fixed to a set of discrete values divides the amplitudes of the Fourier components into the groups corresponding to the energy levels. The Fourier components corresponding to a fixed $A_n$ comprise an invariant oscillation pattern, except for their uniformly varying oscillation amplitudes according to the drive amplitude $E$. In Fig. C1, together with the stabilized mechanical energy, we compare the quadratures $X_c$ and $P_c$ obtained from Eq. (\ref{linearized}) (numerically integrating the equations without resorting to the above formal expansion) with those evolved according to the nonlinear dynamical equations. A good consistency for the results found in the two different ways completely validates the approximation with the base frequency component for the stabilized $X_m(t)$, i.e., the contributions of the twists over the curves in Fig. 2C can be well neglected.

\renewcommand{\thefigure}{D\arabic{figure}}
\renewcommand{\theequation}{D\arabic{equation}}
\renewcommand{\thetable}{D\arabic{table}}
\setcounter{equation}{0}
\setcounter{figure}{0}

\begin{figure}[b!]
\vspace{-0cm}
\centering
\epsfig{file=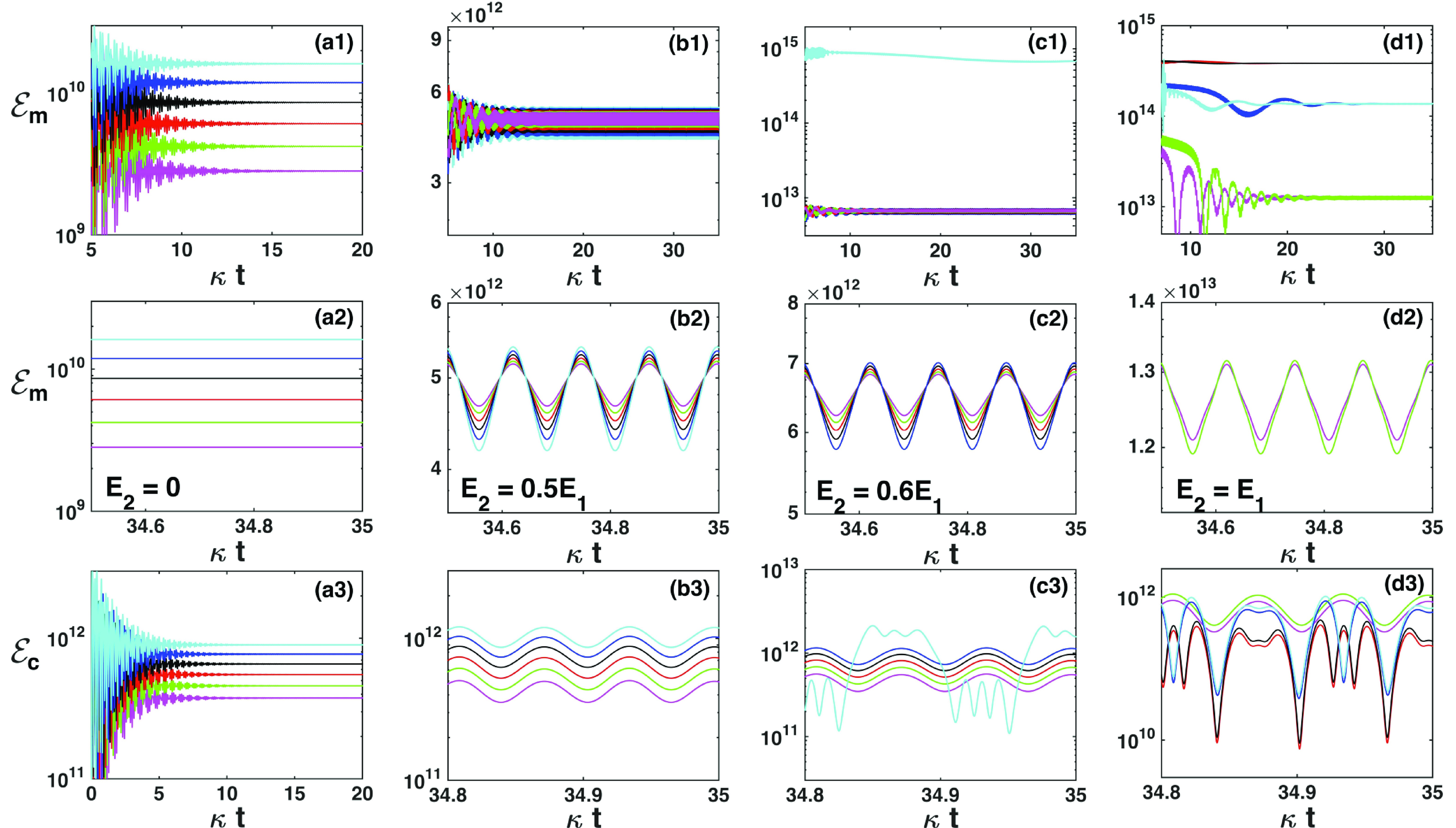,width=1\linewidth,clip=} 
{\vspace{-0.2cm}\caption{{\bf Processes of gradually enhancing 
the resonant field added to the action of a cooling field.} (a1)-(a3) The evolutions of the mechanical energy, the detailed steady states, as well as the associated evolutions of the cavity energy ${\cal E}_c(t)=\frac{1}{2}(X_c^2(t)+P_c^2(t))$, under the sole action of the cooling field. Here six evenly distributed drive amplitudes---$E_1=3.0\times 10^7\kappa$ (pink), $E_1=3.3\times 10^7\kappa$ (green), $E_1=3.6\times 10^7\kappa$ (red), $E_1=3.9\times 10^7\kappa$ (black), $E_1=4.2\times 10^7\kappa$ (indigo), and $E_1=4.5\times 10^7\kappa$ (blue)---are used for the examples. Note that the stabilized results in (a1) and (a2) are not energy levels but the samples among a continuum spectrum of $\langle {\cal E}_m\rangle$. (b1)-(b3) The corresponding results by adding the resonant field with the ratio $E_2=0.5E_1$. (b2) details the energy band shown in (b1). (c1)-(c3) The corresponding results by adding the resonant field with the ratio $E_2=0.6E_1$. (c2) shows the details of the energy band after the one under the highest resonant drive (the blue one) has separated to the upper energy level. (d1)-(d3) The corresponding results by adding the resonant field of $E_2=E_1$. (d2) shows the stabilized oscillations on the level $n=1$.}}
\vspace{-0cm}
\end{figure}

\begin{figure}[h!]
\vspace{-0cm}
\centering
\epsfig{file=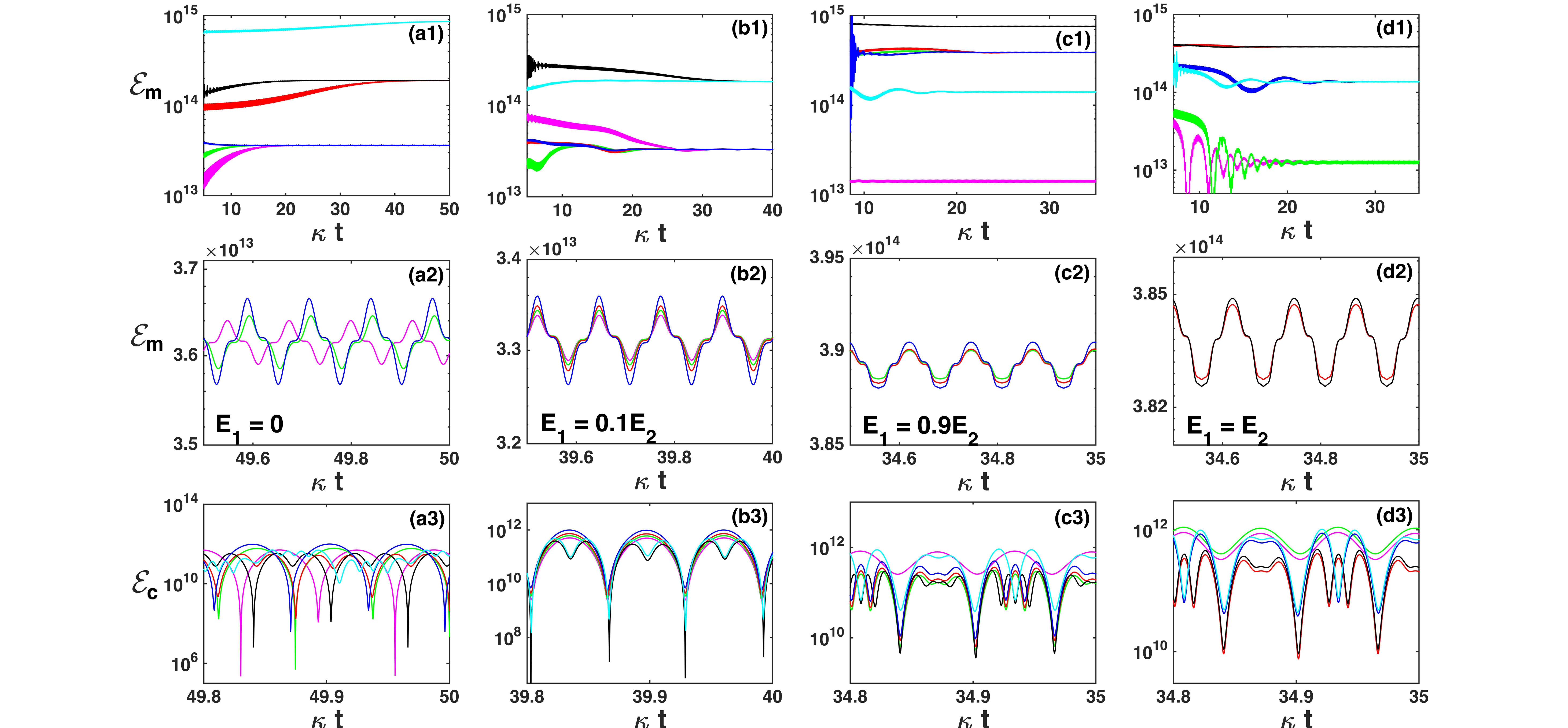,width=1\linewidth,clip=} 
{\vspace{-0.2cm}\caption{{\bf Processes of gradually enhancing the cooling field added to the action of a resonant field.} (a1)-(a3) The evolutions of the mechanical energy 
to different energy levels, the detailed view of the oscillations on the first energy level, as well as the associated evolutions of the cavity energy ${\cal E}_c(t)=\frac{1}{2}(X_c^2(t)+P_c^2(t))$, under the sole action of the resonant field. The drive amplitudes used here are the same as those in Fig. S2---$E_2=3.0\times 10^7\kappa$ (pink), $E_2=3.3\times 10^7\kappa$ (green), $E_2=3.6\times 10^7\kappa$ (red), $E_2=3.9\times 10^7\kappa$ (black), $E_2=4.2\times 10^7\kappa$ (indigo), and $E_2=4.5\times 10^7\kappa$ (blue). (b1)-(b3) The corresponding results by adding the cooling field with the ratio $E_1=0.1E_2$. (c1)-(c3) The corresponding results by adding the cooling field with the ratio $E_1=0.9E_2$. (c2) shows the oscillations on the level $n=3$. (d1)-(d3) The corresponding results by adding the cooling field of $E_1=E_2$. 
(d2) shows the stabilized oscillations on the level $n=3$.}}
\vspace{-0cm}
\end{figure}

\section{Specific effects of two different drives}

Here we adopt the example of optomechanical system in Fig. 1A of the main text to explain the phenomena that can occur in the abstract nonlinear model.
The synchronization and the associated mode locking of optomechanical systems under two cooperating drives can be better understood by starting with only one of the drives and gradually adding up the amplitude of the other. We numerically simulate the processes using Eq. (1) in the main text. In the processes of evolving to the 
levels $n\geq 2$, the existence of various noises \cite{noise-review, book} can affect the evolution courses, but the finally stabilized result will be always on one of the fixed energy levels. We will therefore illustrate the phenomena with the smooth drive amplitudes $E_1=const$ and $E_2=const$ without fluctuations, because the added noisy perturbations to these amplitudes only modify the results without changing the qualitative pictures.

In Fig. D1, we start from acting the cooling field alone. In the absence of the resonant field as in Fig. D1(a1), the stabilized mechanical energy increases with the displayed drive amplitude $E_1$ in a quasi linear way, constituting a continuum spectrum of $\langle{\cal E}_m \rangle$ . The dominant effect in this situation is the exchange of the cavity and mechanical modes or the beam-splitter type coupling of the two modes, 
and then the mechanical oscillator stabilizes quickly under the effective optical damping proportional to the cavity field intensity \cite{oms}. The added resonant field brings about an intensified mechanical displacement and a squeezing type coupling between the cavity and mechanical modes \cite{oms}, so the process from Figs. D1(a1)-D1(a2) to 
Figs. D1(b1)-D1(b2) displays a significant increase of the mechanical energy. Meanwhile, together with the nonlinear saturation and other damping effects, the evolved mechanical energies ${\cal E}_m$ for the different drive amplitudes stabilize in the same range to form something like an energy band, as shown by the numerical simulations in Figs. D1(b1) and D1(b2). The width of the energy band is the difference of the time averages $\langle {\cal E}_m\rangle$ of the stabilized ${\cal E}_m$. With a further strengthened resonant field, the one due to the strongest drive (the blue one) jumps up to somewhere like a energy level; see Fig. D1(c1). Accompanying the jump is the change of the associated cavity energy oscillation pattern in Fig. D1(c3). More energy levels will split out as the resonant field amplitude gets closer to the cooling field amplitude, as illustrated in Fig. D1(d1). The widths of the energy levels (the difference in $\langle {\cal E}_m\rangle$) will also be minimized when the amplitudes of the two different drive fields become equal. 

Since they are dominated by the base frequency component of $\omega_m$, all stabilized mechanical oscillations in Fig. D1 can be in the form
\begin{eqnarray}
X_m(t)=A(E_1, E_2=\chi E_1)\sin(\omega_m t)+d(E_1, E_2=\chi E_1)
\end{eqnarray}
by choosing the proper initial phase, where $0<\chi\leq 1$. On the energy band in Figs. D1(c1) and D1(c2), for example,
the amplitudes $A(E_1, E_2=0.6E_1)$ are almost the same, though the net displacement 
$d(E_1, E_2=0.6E_1)$ increases with the drive amplitude so that there is a considerable width of the energy band. The oscillation amplitude $A(E_1, E_2=0.6E_1)$ for the one on the separate energy level is, however, very different. Such process illustrates a mechanism due to the resonant field---its continuously growing amplitude $E_2$ leads to a uniquely nonlinear response of the mechanical oscillation amplitude $A(E_1, E_2)$, which increases by discrete steps when $E_2$ is sufficiently high. 

On the other hand, one can start from the sole action of the resonant field as in 
Fig. D2. Since it is without optical damping \cite{oms}, the stabilization under resonant field is mainly through more complicated nonlinear saturation. Once the amplitude $E_2$ of the resonant field becomes sufficiently, there is a mechanism as already shown in 
Fig. D1 to leads to the discrete energy levels; see Fig. D2(a1). 
On the energy levels due to a resonant drive alone, however, the mechanical oscillations are not synchronized as shown in the magnified view in Fig. D2(a2), where the oscillation 
\begin{eqnarray}
X_m(t)=A(E_2)\sin\big(\omega_m t+\phi(E_2)\big)+
\text{the higher harmonic components}+d(E_2)
\label{reson}
\end{eqnarray}
caused by different $E_2$ have different phases $\phi(E_2)$. 
An insignificant addition of the cooling field can synchronize these oscillations, as the phenomenon manifests in Fig. D2(b2). In Fig. D2 the positions of the energy levels 
are determined by both resonant field and cooling field; for one example, compare the average positions of the level $n=3$ in Fig. D2(c2) and Fig. D2(d2). Moreover, in the regime where the effects of a strong resonant field dominate, the stabilized amplitude $A(E_1,E_2)$ no longer has a monotonic relation with the amplitude $E_2$; see, for example, the corresponding relation between ${\cal E}_m$ and $E_{2}$ in Fig. D2(a1). 
If $E_2$ varies slightly, the oscillation amplitude $A(E_1, E_2)$ can take transition between those for the discrete energy levels. This is the source of the sensitivity of the energy levels to the drive amplitudes. When the two drive field amplitudes are 
close to each other ($E_1\rightarrow E_2$) and sufficiently high to realize the levels $n\geq 2$, the energy level on which the system is can change even under a slight variation of the drive amplitudes (see Figs. 2B1-2B2 in the main text).  

Despite the complexity in forming the energy levels, their general tendencies with the change of the two different fields are clear. The positions of the energy levels go up with strengthened resonant field, but go down under intensified cooling field. These energy levels will stay on the fixed values when the amplitudes of the two fields are equal to each other, giving rise to the level distribution in Fig. 2A of the main text. It is one consequence of a synchronization process that locks the amplitudes and phases of the mechanical oscillations by the cooling and resonant fields together.

\end{document}